\documentclass[10pt]{aa}
\usepackage{bm}
\usepackage{natbib}
\usepackage{epsfig}
\usepackage{setspace}
\begin{document}
\newcommand{\Ox}{[OI]$\lambda$6300\ }
\newcommand{\He}{HeI$\lambda$5875\ }
\newcommand{\Ha}{H$\alpha$\ }
\newcommand{\kms}{kms$^{\rm -1}$}
\newcommand{\iss}{emission\ }
\newcommand{\Msun}{$M_{\odot}$\ }
\newcommand{\Lsun}{$L_{\odot}$\ }
\newcommand{\A}{$A_{v}$\ }
\newcommand{\T}{$T_{eff}$\ }
\newcommand{\M}{$M_{bd}$\ }
\newcommand{\MJ}{$M_{J}$\ }
\newcommand{\Mstar}{$M_{\star}$\ }
\newcommand{\Lstar}{$L_{\star}$\ }
\newcommand{\OI}{[OI]$\lambda\lambda$6300,6363\ }
\twocolumn
\begin{center}
\LARGE
{{\bf A resolved outflow of matter from a brown dwarf}}\\
\end{center}

\begin{flushleft}
\normalsize
{{\bf Emma T. Whelan$^{1}$, Thomas P. Ray$^{1}$, Francesca Bacciotti$^{2}$,  
Antonella Natta$^{2}$, Leonardo Testi$^{2}$ \& Sofia Randich$^{2}$}}

\scriptsize

$^{1}$ Dublin Institute for Advanced Studies, 5 Merrion Square, Dublin 2, 
Ireland \\
$^{2}$ Osservatorio Astrofisico di Arcetri, INAF, Largo E.Fermi 5, I-50125 Firenze, Italy
\end{flushleft}

{\bf The birth of stars involves not only accretion but also, counter-intuitively, the expulsion of matter in the form of highly supersonic outflows$^{1,2}$. Although this phenomenon has been seen in young stars, a fundamental question is whether it also occurs amongst newborn brown dwarfs: these are the so-called 'failed stars', with masses between stars and planets, that never manage to reach temperatures high enough for normal hydrogen fusion to occur$^{3}$. Recently, evidence for accretion in young brown dwarfs has mounted$^{4-6}$, and their spectra show lines that are suggestive of outflows$^{7-9}$. Here we report spectro-astrometric data that spatially resolve an outflow from a brown dwarf. The outflow's characteristics appear similar to, but on a smaller scale than, outflows from normal young stars. This result suggests that the outflow mechanism is universal, and perhaps relevant even to the formation of planets.}

The nearby $\rho$-Ophiuchi Cloud (at a distance of 125 pc; ref. 10) is an 
excellent example of a stellar nursery. It was as part of a near-infrared survey
of this region$^{11}$ that $\rho$-Oph 102 was first detected. Its mass is given 
as 60\MJ (ref.7), where 1\MJ equals the mass of Jupiter or $0.001 M_{\odot}$. This places 
it firmly within the brown dwarf mass range 
(\bm{$0.013 M_{\odot} \leq  M \leq 0.075 M_{\odot}$})$^{3}$ 
and indeed this object has been spectroscopically 
confirmed to be a BD$^{7}$. There is also strong evidence for the presence of an 
accretion disk$^{4}$ and, in fact, the accretion rate (derived from 
its H$\alpha$ profile)$^{7}$ is estimated to be approximately 
10$^{-9}$ \Msun yr$^{-1}$. Finally it has been noted that its spectrum 
contains a number of forbidden emission lines suggesting an outflow$^{7}$. 

Tracing an outflow from a young star is done using a variety of techniques depending
upon wavelength. When the star itself is sufficiently evolved to be optically visible, 
i.e.\ much of the surrounding natal gas and dust has been driven away, 
the outflow can be followed almost right back to the star through its 
permitted and forbidden line emission$^{1}$. 
Nearest the source, the light from the collimated outflow is most intense, giving rise to 
the so-called ``micro-jets'' observed, for example, from T Tauri stars, the 
precursors to stars like our sun$^{12}$. 
In extreme cases, the outflow is observed only very close
($\leq$1\arcsec) to the source$^{13}$, i.e.\ within the typical ``seeing'' disc for 
ground based telescopes. It is under such circumstances that the technique 
of spectro-astrometry comes to the fore (see Methods and Supplementary Information). 

Our high resolution echelle observations (see Methods and Fig.\ 1) of $\rho$-Oph 102 
first of all show its forbidden emission lines, e.g.\ [OI]$\lambda\lambda$6300,6363
and [SII]$\lambda$6731,
are moderately blueshifted (V$_{r}\approx$ -45\,\kms). Such radial velocities are very 
similar to those seen in outflows from classical T Tauri stars, 
and might be expected of an outflow from a young BD, as its escape velocity 
is similar$^{14}$. We note also the absence of a redshifted 
outflow component. This is again typical of T Tauri stars and is interpreted in terms of 
an obscuring disk that hides the redshifted component from our view$^{13}$. We can thus 
immediately infer the presence of a disk around this brown dwarf (see below for a limit on its 
size). Such a conclusion is also in line with the mid-infrared excess seen from this 
object$^{4}$. Note that all velocities are with respect to the systemic velocity of the 
BD. The latter was derived from the Li\,$\lambda$6708 photospheric absorption line and 
equals 7$\pm$8 \kms\ in the Local Standard of Rest frame.  

Another indication that an outflow is present comes from a cursory examination of the 
H$\alpha$ line (Fig.\ 2). Its profile is clearly asymmetrical, i.e.\ the blueshifted
wing of the line appears to be absorbed in a P-Cygni like fashion. We note however 
that we do not observe a classical P-Cygni profile, i.e\ one that dips below 
the continuum. Such a profile is in any event a rare occurrence even amongst T Tauri 
stars$^{15}$.       

If we are dealing with a scaled-down version of the outflow phenomenon seen in 
T Tauri stars then we expect the centre of emission in forbidden lines, 
to be spatially offset from the continuum, i.e.\ the BD. 
This offset is due to the fact that, for a collimated outflow, such lines are quenched 
close to the star once the electron density becomes high enough$^{13}$. In the case 
of outflows from T~Tauri stars, typical offsets of 30-75 AU (0$\farcs$2-0$\farcs$5 
at 150 parsecs) are seen, for example, in [OI]$\lambda$$\lambda$6300,6363 and the 
red [SII] doublet$^{13}$. If we assume brown dwarf outflows have similar opening angles and 
velocities to those from T~Tauri stars, then the point at which the critical density 
is reached, na\"{i}vely scales with \.M$_{jet}^{0.5}$ where \.M$_{jet}$ is the jet's 
mass loss rate (see Supplementary Information). Assuming the latter depends linearly on the accretion rate, 
we would expect typical spatial offsets to be 3-10 times smaller in brown dwarf outflows in 
comparison to those from T Tauri stars.

Spectro-astrometric (emission centroid offset versus velocity) plots are shown in 
Fig.\ 1 for the [OI]$\lambda$6300, [OI]$\lambda$6363, H$\alpha$ 
and [SII]$\lambda$6731 lines. Because of high electron densities close to the BD, 
[SII]$\lambda$6717 was too weak to provide a usable spectro-astrometric signal. Here the 
spatial offsets were measured after continuum subtraction (see the Methods and Supplemenatry Information for details).

We now consider the main results from various offset versus velocity plots. First, the centroids of {\em all} the measurable forbidden lines are displaced to the 
south, i.e.\ have negative offsets with respect to the continuum.  These offsets reach a 
maximum, of 0$\farcs$08$-$0$\farcs$1, 
at a blue-shifted velocity of approximately -40\,\kms. We have already noted the absence 
of any corresponding redshifted emission and that this suggests the presence of an 
obscuring disk. The scale of the blueshifted offset would suggest a minimum (projected) 
disk radius of 0$\farcs$1 ($\geq$ 15\,AU at the distance of the $\rho$ 
Ophiuchi Cloud) in order to hide any redshifted component.

Second, there is no clear spatial offset in H$\alpha$ even though its higher signal to noise 
potentially allows us to measure even smaller offsets than observed in the forbidden 
lines. This is in agreement with the idea that most of the H$\alpha$ emission arises 
from accretion$^{7}$ on much smaller scales than are being probed here. It is also worth 
pointing out that if, as is almost certainly the case, the blue-ward dip in the H$\alpha$ 
profile is caused by P Cygni-like absorption from an outflowing wind in front of the star, 
no offset should be expected. 

Third, both the line profiles and the spectro-astrometric signatures are very similar (albeit 
on somewhat smaller scales) to what is seen in T Tauri stars with ``micro-jets''. 
In particular the observed velocities and offsets in the various forbidden lines 
are within the range we would expect for a collimated 
outflow from a brown dwarf (see also Supplementary Information). We suggest that direct imaging, (using, for example, Fabry-Perot systems) of this and 
other candidate brown dwarf outflows, should now be attempted. Such observations however will 
be very challenging, even with large telescopes, because of the expected faintness of 
the outflow.   

\section*{Methods}

\subsection*{Spectro-astrometry}

Conceptually the principles of spectro-astrometry are easy to understand. The profile of a star is smeared by atmospheric turbulence to appear gaussian (at least to a first approximation) rather than point-like. Whereas the width of the profile is determined by the so-called seeing, how accurately we can determine the centroid of emission is, in theory for fixed seeing, limited only by the strength of the observed signal to noise ratio. Increasing the total number of detected photons increases the positional, or astrometric, accuracy, so that, in principle, milliarcsecond precision is possible with very large ground based telescopes$^{16-18}$.

Consider now a long-slit spectrum of a close binary system consisting of two virtually identical stars. We will assume that the slit is orientated along the same position angle as the binary. (We note that strictly this is not necessary: it is only necessary that the slit is not orthogonal). If the separation of the binary is considerably less than the seeing, the profile of the system in the spatial direction will consist of a single gaussian-that is, the system is unresolved and the centroid of emission will lie exactly between the two components. Now suppose that one of the two stars differs slightly from the other in being a strong H$\alpha$ emitter; in such a case, the emission centroid will shift towards that star in the spectrum at the position of the H$\alpha$ line. In this way it is possible to resolve certain types of binaries with separations well within the seeing limit$^{19}$. In the case of a jet (pure emission line region) plus star (continuum source), one can go further and interpolate the continuum across a line, thereby allowing its contribution to be removed. It is then possible to measure separately the spatial centroid of the pure emission line region and determine its offset with respect to the continuum, that is, the parent star. Moreover, as the line can be emitted over a range of wavelengths, owing to the Doppler effect, it may also be feasible to recover spatio-kinematic information. For example, if the jet is bipolar, that is, it has oppositely directed blue- and redshifted flows from the source; the emission centroid of the red and blue wings of the line will be displaced to opposite sides of the continuum centre.

The detailed method by which we measure offsets can briefly be described as follows. First, the centroid of the continuum emission in the spatial direction is determined using a one-dimensional gaussian fit. The line of such centroids, in the dispersion direction but excluding any region where emission lines are present, is then fitted with a second-order polynomial, over a range of typically 200-300 Å. In this way, instrumental curvature and tilting, with a characteristic frequency many times larger than the width of any line, is determined. The fit, to the centre of the continuum, is then subtracted from the actual measured centroids, leaving residuals that are evenly scattered about the abscissa (that is, the fit defines the zero offset line). The continuum data points shown in Fig. 1 are thus the residuals. Finally, the two-dimensional fit to the continuum, broadened to take account of the point spread function, is subtracted from the emission lines. Any emission line offsets are then measured.

The accuracy (in arcseconds) of the method is set by the error in the centroid of the gaussian fit, which depends on the seeing and the number of detected photons, {\it N}. Formally, the error is given by Seeing/[2(2 ln 2)1/2N1/2], assuming that photon noise is the only source of noise. {\it N}, of course, is a function of the binning and the spatial sampling (pixel width). This explains why, for example, we can achieve a higher spectro-astrometric accuracy with a bright line, such as H$\alpha$ than a weak one, for example, the [SII]$\lambda$6731 line. In some cases, it is necessary to bin up a weak line in the dispersion direction, as we have done to varying degrees for the [O I] doublet and the [SII]$\lambda$6731 line, to achieve sufficient signal to noise ratio. Note that we sometimes use different binning factors for the continuum, in comparison with the line, so as to achieve a similar signal to noise ratio in both. This allows us to have comparable offset errors in both components, and to define the common 1$\sigma$ error lines shown in Fig. 1. As can be seen from the plots, the typical limiting offset that we can measure in the spatial direction (3$\sigma$) is around 30 mas. This corresponds to 4.5 au at the distance to the $\rho$ Ophiuchi cloud.

\subsection*{Echelle spectroscopy}
The high resolution spectra of $\rho$ Oph 102 were taken with the UV-visual Echelle Spectrograph (UVES) on the European Southern Observatory's 8 m Kueyen Telescope, one of the telescopes in the Very Large Telescope (VLT) suite, in May 2003. A total of three 45 min exposures of the target were made, together with a series of flats and biases as well as an observation of an arc lamp for wavelength calibration. The slit was orientated north-south and had a width of 1 \arcsec while the seeing was 0\farcs65. The central wavelength was set at 580 nm, giving a spectral range of 450-680 nm. Only the red part of the spectrum from 580 to 680 nm, however, was analysed. The pixel scale was 0\farcs182 and the spectral resolution R=40,000. The data were reduced using standard Image Reduction and Analysis Facility (IRAF) routines.

\begin{enumerate}

\item{Eisl\"offel, J., Mundt, R., Ray, T.~P., \& Rodr\'{\i}guez, L.~F.\ 
Collimation and propagation of stellar jets. {\it Protostars and Planets IV, 
University of Arizona Press.} 815-840 (2000).} 

\item{K\"onigl, A.~\& Pudritz, R.~E.\ Disk winds and the accretion-outflow connection. 
{\it Protostars and Planets IV, University of Arizona Press.} 759-787 (2000).}
 
\item{Basri, G.\ Observations of brown dwarfs. 
{\it Ann.\ Rev.\ Astron.\ Astrophys.} {\bf 38}, 485-519 (2000).}

\item{Natta, A., Testi, L., Comer{\' o}n, F., Oliva, E., D'Antona, F., Baffa, C., 
Comoretto, G., \& Gennari, S.\ Exploring brown dwarf disks in rho Ophiuchi. 
{\it Astron.\ Astrophys.}, {\bf 393}, 597-609 (2002).}

\item{Pascucci, I., Apai, D., Henning, T., \& Dullemond, C.~P.\ The first detailed look 
at a brown dwarf disk. {\it Astrophys.\ J.}, {\bf 590}, L111-L114 (2003).} 

\item{Jayawardhana, R., Mohanty, S., \& Basri, G.\ Evidence for a T Tauri phase in young 
brown dwarfs. {\it Astrophys.\ J.} {\bf 592}, 282-287 (2003).} 

\item{Natta, A., Testi, L., Muzerolle, J., Randich,S., Comer{\'o}n, F., \& Persi, P.\ 
Accretion in brown dwarfs: An infrared view. {\it Astron.\ Astrophys.} 
{\bf 424} 603-612 (2004).} 

\item{Comer{\' o}n, F., Fern{\' a}ndez, M., Baraffe, I., Neuh{\" a}user, R., \& 
Kaas, A.~A.\ New low-mass members of the Lupus 3 dark cloud: Further indications of 
pre-main-sequence evolution strongly affected by accretion. {\it Astron.\ Astrophys.} 
{\bf 406}, 1001-1017 (2003).} 

\item{Fern{\' a}ndez, M.~\& Comer{\' o}n, F.\ Intense accretion and mass loss of a very 
low mass young stellar object. {\it Astron.\ Astrophys.} {\bf 380}, 264-276 (2001).}

\item{Greene, T.~P.~\& Young, E.~T.\ Near-infrared observations of young stellar objects 
in the rho Ophiuchi dark cloud. {\it Astrophys.\ J.}, {\bf 395}, 516-528 (1992).} 

\item{de Geus, E.~J., de Zeeuw, P.~T., \& Lub, J.\ Physical parameters of stars in 
the Scorpio-Centaurus OB association. {\it Astron.\ Astrophys.} {\bf 216}, 44-61 (1989).} 

\item{Dougados, C., Cabrit, S., Lavalley, C., \& M{\' e}nard, F.\ T Tauri stars microjets 
resolved by adaptive optics. {\it Astron.\ Astrophys.} {\bf 357}, L61-L64 (2000).}

\item{Hirth, G.~A., Mundt, R., \& Solf, J.\  Spatial and kinematic properties of the 
forbidden emission line region of T Tauri stars. {\it Astron.\ Astrophys.\ Suppl.} 126, 
437-469 (1997).}

\item{Masciadri, E., \& Raga, A.~C.\ Looking for outflows from brown dwarfs. 
{\it Astrophys.\ J.}, {\bf 615}, 850-854 (2004).} 

\item{Muzerolle, J., Calvet, N., \& Hartmann, L.\ Emission-line diagnostics of T Tauri 
magnetospheric accretion. II. Improved model tests and insights into accretion physics. 
{\it Astrophys.\ J.} {\bf 550}, 944-961 (2001).}

\item{Takami, M., Bailey, J., Gledhill, T.~M., Chrysostomou, A., \& 
Hough, J.~H.\ Circumstellar structure of RU Lupi down to AU scales. 
{\it Mon.\ Not.\ R.\ Astron.\ Soc.} {\bf 323} 177-187 (2001).}

\item{Whelan, E.~T., Ray, T.~P., \& Davis, C.~J.\ 
Paschen beta emission as a tracer of outflow activity from T-Tauri stars, as compared to 
optical forbidden emission. {\it Astron.\ Astrophys.} {\bf 417} 247-261 (2004).}

\item{Takami, M., Bailey, J., \& Chrysostomou, A.\ 
A spectro-astrometric study of southern pre-main sequence stars. Binaries, outflows, and 
disc structure down to AU scales. {\it Astron.\ Astrophys.} {\bf 397} 675-691 (2003).} 

\item{Bailey, J.\ Detection of pre-main-sequence binaries using spectro-astrometry. 
{\it Mon.\ Not.\ R.\ Astron.\ Soc.} {\bf 301} 161-167 (1998).} 

\end{enumerate}

{\bf Supplementary Information} is linked to the online version of the paper at www.nature.com/nature but requests for supplementary materials can also be addressed to ewhelan@cp.dias.ie 

{\bf Acknowledgements} This work was supported in part by Science Foundation Ireland and the JETSET Marie Curie reserach training network.

\newpage

\begin{figure*}
\epsfig{file= 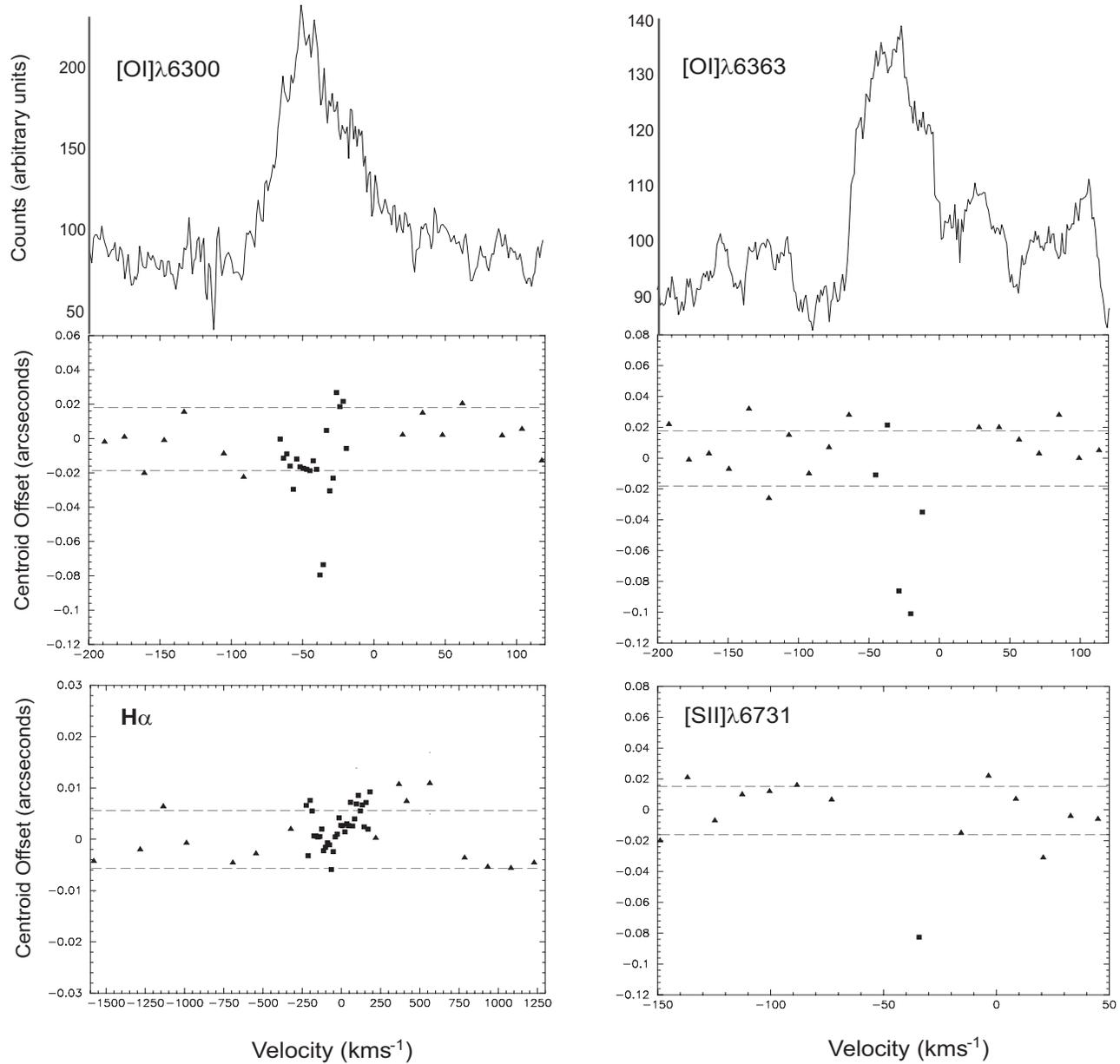, width=17cm, height=17.5cm}
\caption{Line profiles (top row) and spectro-astrometric plots (middle row) for the 
[OI]$\lambda$6300 and [OI]$\lambda$6363 doublet and spectro-astrometric plots (bottom row) for the \Ha and [SII]$\lambda$6731 lines. The much narrower 
corresponding night-sky lines, peaking around -9\kms, have been subtracted although 
no offset measurements were made in their vicinity (thus accounting for the data point 
gaps). Continuum and line offset points are represented by black triangles and squares 
respectively. All velocities are systemic and spatial offsets are in the north-south 
direction (in arcseconds) with negative offsets to the south. Dashed lines delineate 
the $\pm 1\sigma$ error envelope. For \Ha, note the much smaller offset scale. The [SII] line is blue-shifted to 
around -40\kms.}
\end{figure*}

\newpage

\begin{figure*}
\epsfig{file = 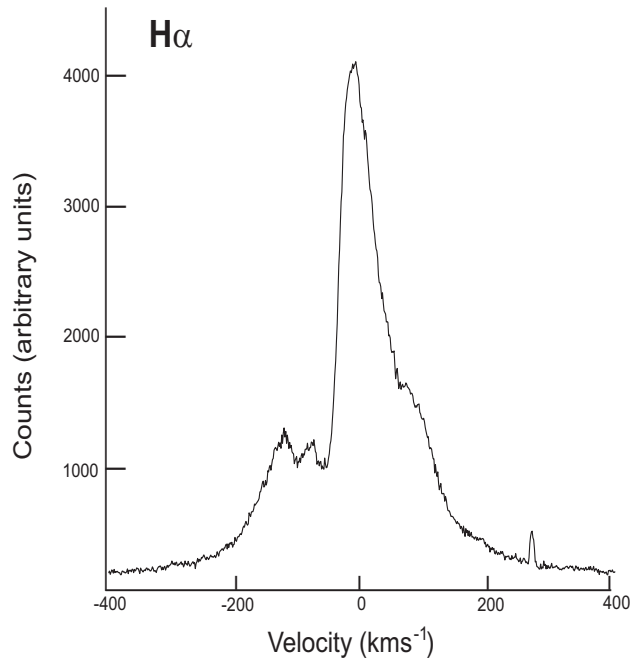, width=10cm, height=11cm}
\caption{{\bf The outflow signature in the H$\alpha$ profile of $\rho$-Oph~102}. The P Cygni-like dip in the line profile is a strong signature of outflow activity. \Ha emission from the brown dwarf is absorbed as it passes through material moving outwards along our line of sight. Because this material is moving towards us, the dip is on the blueward side of the line. Classical T Tauri stars are strong \Ha emitters, and P Cygni \Ha profiles originally confirmed that such protostars drive outflows. The dip in the \Ha emission of $\rho$ Oph 102 is at approximately the outflow radial velocity determined from the forbidden lines.}

\end{figure*}

\end{document}